\documentclass{INTERSPEECH2023}

\usepackage[colorinlistoftodos]{todonotes}
\usepackage{subcaption}
\usepackage[utf8]{inputenc}
\usepackage{float}
\usepackage{multirow}
\usepackage{epstopdf}
\interspeechcameraready 
\title{Adaptation of Tongue Ultrasound-Based Silent Speech Interfaces \protect\\Using Spatial Transformer Networks}
\name{\begin{tabular}{c}L\'aszl\'o T\'oth$^1$, Amin Honarmandi Shandiz$^1$, G\'abor Gosztolya$^{1,2}$, Tam\'as G\'abor Csap\'o$^{3}$\end{tabular}
}
\address{
  $^1$Institute of Informatics, University of Szeged, Hungary\\
  $^2$ELRN-SZTE Research Group on Artificial Intelligence, Szeged, Hungary \\
  $^3$Department of Telecommunications and Media Informatics, \\
   Budapest University of Technology and Economics, Budapest, Hungary }
\email{\{tothl, shandiz, ggabor\}@inf.u-szeged.hu, csapot@tmit.bme.hu}

\begin{document}

\maketitle
\begin{abstract}
Thanks to the latest deep learning algorithms, silent speech interfaces (SSI) are now able to synthesize intelligible speech from articulatory movement data under certain conditions. However, the resulting models are rather speaker-specific, making a quick switch between users troublesome. Even for the same speaker, these models perform poorly cross-session, i.e. after dismounting and re-mounting the recording equipment. To aid quick speaker and session adaptation of ultrasound tongue imaging-based SSI models, we extend our deep networks with a spatial transformer network (STN) module, capable of performing an affine transformation on the input images. Although the STN part takes up only about 10\% of the network, our experiments show that adapting just the STN module might allow to reduce MSE by 88\% on the average, compared to retraining the whole network. The improvement is even larger (around 92\%) when adapting the network to different recording sessions from the same speaker.

 \end{abstract}

\section{Introduction}

The goal of articulatory-to-acoustic mapping (AAM) is to synthesize speech from data capturing the movement of the articulatory organs. This is the purpose of the giving the background to the development of 'Silent Speech Interface' systems (SSI~\cite{Denby2010,Gonzalez-Lopez2020}). Ideally, these interfaces would record the articulation and synthesize speech based on the movement of the organs -- without the user of the device actually producing any sound. 
The typical input of AAM can be a video of the lip movements~\cite{Hueber2010,Ephrat2017,Mira2021,Oneata2021,Saleem2022,Wang2022}, ultrasound tongue imaging (UTI)~\cite{Hueber2010,Denby2004,Csapo2017c,Tatulli2017,Moliner2019,Csapo2019a,SottoVoce,Csapo2020c,Zainko2021,Toth-is2021}, or several other modalities (e.g., MRI, EMA, PMA, EOS, radar, multimodal, etc.). All of the articulatory tracking devices are highly sensitive to 1) the alignment of the recording equipment across sessions, 2) the actual speaker's anatomy. For example, in the case of ultrasound recordings, the probe fixing headset has to be remounted onto the speaker for each recording session. This inevitably causes the recorded ultrasound videos to become misaligned between each recording session~\cite{Csapo2020d}. Moreover, there are large individual differences across speakers, so even a system trained on the data of several speakers may still perform poorly for a new speaker.

There have already been several cross-session and cross-speaker studies, of which we mention only those related to imaging. To handle the session dependency of UTI-based synthesis, Gosztolya et al. used data from different sessions~\cite{Gosztolya2020}. Ribeiro et al. reported that, for a speaker-in\-de\-pen\-dent system, unsupervised model adaptation can improve the results for silent speech~\cite{Ribeiro2021}. In a multi-speaker framework, Shandiz et al. experimented with 
x-vectors features extracted from speakers, leading to a marginal improvement in the spectral estimation step~\cite{Toth-is2021}. Zhang et al. evaluated UTI and lip video based unconstrained multi-speaker voice recovery with a transfer learning strategy and encoder-decoder architecture~\cite{Zhang2021b}. There have been more studies on multi-speaker lip-to-speech synthesis~\cite{Mira2021,Oneata2021,Saleem2022,Wang2022}. One of the first papers that could produce intelligible speech for unseen speakers was based on WGAN with new additional critics and losses~\cite{Mira2021}. Another study proposed speaker disentanglement by inputting speaker identities or embeddings to the DNN~\cite{Oneata2021}. An end-to-end ResNet-based model was claimed to outscore previous approaches on unseen speakers~\cite{Saleem2022}, while Wang et al. applied cross-modal knowledge transfer and voice conversion to generate good quality speech with high naturalness~\cite{Wang2022}.

Most of the above approaches hope to solve speaker sensitivity simply by acquiring articulatory training data from a large quantity of speakers. In this study, we experiment with a direct adaptation of an UTI-based SSI network to the actual speaker or session. To avoid the need for a full retraining, we extend our network with a spatial transformer network (STN) module and retrain only this module during the adaptation step. The STN learns an affine transformation on the input images, and our assumption is that this transformation should mostly be able to compensate for the misalignment of the recording device, and to a certain extent also for the inter-speaker differences.

\section{The UTI-to-Speech framework}

\begin{figure}[!t]
\centering
\includegraphics[width=0.43\textwidth]{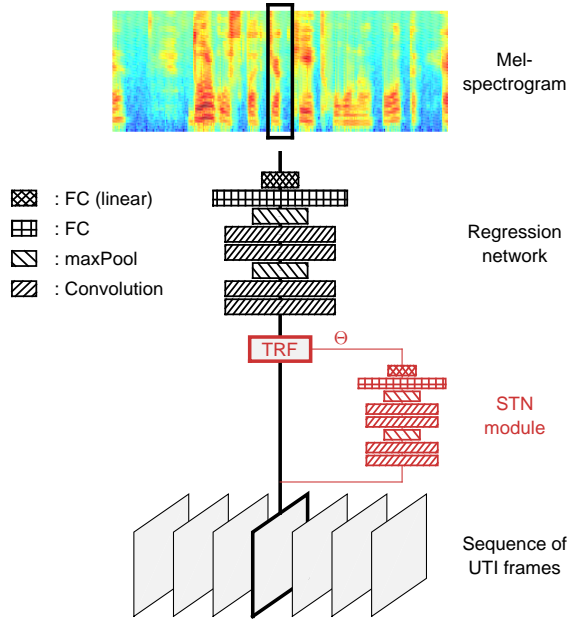}
\caption{\textit{Illustration of the network architecture used, where {FC} denotes the Fully-Connected layer.}} \label{fig:stn}
\vspace{-20pt} 
\end{figure}

Approaches for articulatory-to-acoustic mapping similar to ours were described several times~\cite{Csapo2020c,Toth-is2021,Yu-2021},
so we just give a brief overview here (see also Fig.~\ref{fig:stn}). The input to our system is a sequence of ultrasound tongue images, which record the movement of the articulators at a relatively high frame rate. Our goal is to estimate the speech signal produced during articulation, but instead of producing a speech signal, we estimate only a mel-spectrogram at the network output, and we apply a neural vocoder to convert the spectrogram to speech. The main advantage of this approach is that the mel-spectrum is a very dense representation, which is easier to estimate than the speech signal itself, and we can apply large pre-trained networks for the synthesis step, such as WaveGlow~\cite{waveglow}. Our training data consists of precisely time-aligned pairs of ultrasound videos and speech signals, so the ultrasound-to-spectrogram mapping can be performed on a simple frame-by-frame basis, converting each ultrasound image to a spectral vector~\cite{Tatulli2017, Csapo2020c}. While this simple arrangement already performs reasonably well, significant improvement can be achieved by involving the input context, that is, by using a block of video frames as input instead of just one image. Several network architectures have been proposed to process 3D blocks of input data, for video processing in general~\cite{Ji2013, Tran, feichtenhofer2019slowfast}, and for ultrasound input in particular~\cite{Moliner2019, SottoVoce, toth20203d, Zainko2021, Saha-ultra2speech}. In the experimental section we will experiment both with 2D and 3D Convolutional Neural Networks (CNNs) for the mapping task.
The problem could also be addressed even in the lack of aligned training data using encoder-decoder networks~\cite{Ribeiro2021, Zhang2021b} or video transformers~\cite{bertasius2021space,serdyuk2022transformer}.  

\section{The Spatial Transformer Network}

The concept of the Spatial Transformer Network (STN) was motivated by the fact that in image recognition the actual input is frequently shifted, rotated or scaled compared to the training data, due to changes in camera angle, distance, and other factors~\cite{STN}. Although CNNs are somewhat invariant to translations, their overall flexibility would greatly improve by introducing a dynamic mechanism that can spatially transform the input image by an appropriate transformation before classification. The STN is a network module that can be inserted into a CNN architecture at any point, but typically it is applied between the input and the first network layer. In this arrangement, the STN performs an affine transformation on the input image, and returns a transformed image of the same size. As it ma\-ni\-pu\-lates only the input, it can be combined with practically any type of classification or regression network. The affine transform defined by the STN is quite powerful and includes translation, scaling, rotation, shearing and cropping as special cases. 

The STN consists of three main parts, namely the localization network, the grid generator and the sampler~\cite{STN}. The grid generator and the sampler together are responsible for executing the affine transformation defined by the parameter $\theta$, which consists of 6 components in the case of a 2D input~(see Fig.~\ref{fig:stn}). 

The grid generator and the sampler are differentiable. This is vital for propagating the error back to the third component, the localization network, which learns to estimate the $\theta$ parameters, conditioned on the actual input image. The localization network can take any form, but its uppermost layer must be a regression layer to produce the $\theta$ values.

While we explained the STN concept assuming 2D images, the whole idea can be naturally extended to 3D data blocks~\cite{STN, Bas_2017_ICCV}. The 3D variant has already been used in visual speech recognition (e.g., lip reading) by Yu and Wang~\cite{VisualSpeech-STN}. We discuss the options for a 3D extension in Section~\ref{Conf3D}.

Although the original concept uses the STN to decrease the variance of the training data by transforming the input images to a canonical, expected pose, it may also be useful in handling a domain mismatch between the training and testing conditions. It was applied for domain adaptation in various image processing situations~\cite{STN-domainAdaptation, STN-domainShift}, and here we evaluate its efficiency for speaker and session adaptation in UTI-to-speech conversion.

\begin{figure}[!t]
\centering
\includegraphics[width=0.43\textwidth]{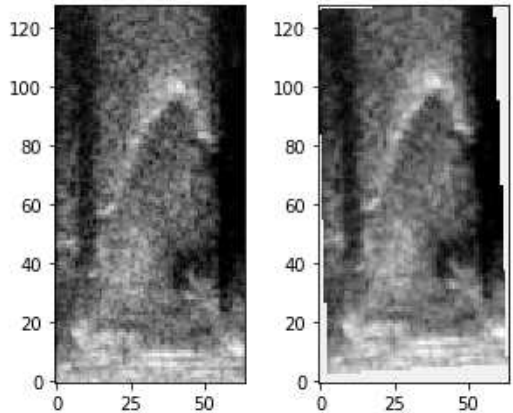}
\caption{\textit{An example UTI image, before(left) and after(right) STN.}} \label{fig:us}
\vspace{-20pt} 
\end{figure}

\section{Experimental Set-Up}

\subsection{Data Acquisition and Preprocessing}

Several Hungarian male and female subjects with normal speaking abilities were recorded while reading sentences aloud (altogether 209 sentences each), of which two female and two male speakers were selected for the speaker adaptation experiments (with speaker IDs 048, 049, 102 and 103). The average duration of the recordings was about 15 minutes per speaker. 
For the cross-session experiments one female speaker from the 4 speakers (speaker 048) was asked to record 4 additional sessions on different days (obviously, with dismounting and remounting the UTI recording device between the sessions). These 4 additional sessions are shorter than the main session of the given speaker, with an average duration of 3.5 minutes. All sessions were divided into training, validation and test partitions randomly, in a 85-10-5 ratio. The speech signal and the ultrasound articulatory data were recorded in parallel, and were synchronized using the software provided with the UTI equipment, Figure~\ref{fig:udi}. The tongue movement was recorded in a midsagittal orientation using the “Micro” ultrasound system of Articulate Instruments Ltd. at 81.67 fps. The speech signal was recorded with an 
omnidirectional condenser microphone. For more details on the recording set-up see~\cite{Csapo2017c}. In the experiments we used the raw scanline data of the ultrasound as the training input, after being resized to 64×128 pixels. The intensity range of the images was min-max normalized to the [-1, 1] interval.
The speech signals were converted to a mel-spectrogram with 80 spectral components, according to the requirements of WaveGlow. These 80-component spectral vectors served as the training targets for our neural networks, after standardization to zero mean and unit variance.

\subsection{Network configuration}
\label{Conf}

In the first set of experiments we applied a simple 2D convolutional (2D-CNN) network that transforms one ultrasound image to one spectral vector, and the localization network of the STN also was a 2D-CNN. Although this approach is suboptimal, it may already show whether the STN is suitable to perform domain adaptation, while visualization and interpretation is easier in 2D. Fig.~\ref{fig:stn} illustrates this network arrangement, while Fig.~\ref{fig:us} demonstrates the effect of STN on an actual ultrasound image. 
The 2D-CNN network had a quite simple and traditional structure. It consisted of 4 convolutional layers with 30-60-90-120 filters, with a MaxPooling layer after every second layer. The convolutional processing was followed by a fully connected layer of 300 neurons, and the linear output layer of 80 neurons. All hidden layers applied the Swish nonlinearity, and overfitting was minimized by placing dropout layers (p=0.2) after each processing layer. We applied the mean squared error (MSE) loss function, which was minimized using the Adam optimizer with a batch size of 100 and a learning rate of $2\cdot10^{-3}$. Training was halted using early stopping on the validation set.

The input of the STN module is the same image as that of the spectral regression module, and it also has to perform regression. Hence we used exactly the same 2D-CNN architecture for the STN as for the main network. However, as the STN has to estimate only 6 parameters (the $\theta$ vector) instead of 80, we configured it to be much smaller than the regression network. The number of free parameters in the STN module was only about 10\% of that of the whole network.

\subsection{Extension to 3D}
\label{Conf3D}

The simple framework presented above can be significantly improved by extending the input from just one image to a sequence of video frames. Such 3D blocks of input data can be processed by various network types, such as by combining 2D-CNN and LSTM layers, by 3D-CNN models, or by using Convolutional LSTM layers~\cite{Moliner2019, SottoVoce, Saha-ultra2speech, toth20203d, Predicting-ConvLSTM}. Here, we applied the 3D-CNN architecture introduced
by T\'oth et al.~\cite{toth20203d, Toth-is2021}.
This network had the same global architecture as the 2D-CNN presented above. However, the convolutions were extended to the time axis as well, in a somewhat special arrangement. The first Conv3D layer applied a large stride along the time axis, dividing the sequence of 25 input images into 5 five blocks along time. The next two Conv3D layers had a filter size of 1 along time, practically processing the 5 blocks separately. Finally, the extracted information was fused along time by the last Conv3D layer and by the topmost two fully connected layers.

As regards the STN, there are several options to extend it to 3D. First, we could reformulate the affine transformation to operate on 3D data. However, for our ultrasound data, translation, shearing and rotation along the time axis seemed to be unnecessary. Hence, we chose to keep the transformation in 2D. The second question was whether the localization network should operate on 2D or 3D data. 
We chose the technically simpler solution, and used the same 2D localization network as for the 2D-CNN. In this configuration, the STN applies the same 2D transformation to all the 25 images of the actual input block.

\section{Results and discussion}
\begin{table}[ht]
\caption{MSE rates of the 2D-CNN on the dev set of the 4 speakers (top), and for the 4 extra sessions of speaker 048 (down).} \label{tab:baseline}
\centering
\renewcommand{\arraystretch}{1.1} 
\begin{tabular}{|l||c|c|c|c|}
\hline
~ & s048 & s049 & s102 & s103\\
\hline
\hline
~no STN~ & 0.361 & 0.387 & 0.390 & 0.343\\
\hline
~with STN~ & 0.358 & 0.396 & 0.389 & 0.340\\
\hline
\end{tabular}
\vspace{1mm}

\centering
\renewcommand{\arraystretch}{1.1} 
\begin{tabular}{|l||c|c|c|c|}
\hline
& s048-2 & s048-3 & s048-4 & s048-5\\
\hline
\hline
~no STN~ & 0.505 & 0.462 & 0.464 & 0.504\\
\hline
~with STN~ & 0.495 & 0.458 & 0.467 & 0.483\\
\hline
\end{tabular}
\vspace{-5mm}
\end{table}
First of all, we examined whether the inclusion of the STN module has any positive effect on the results when using only training data from one speaker and one session. Table~\ref{tab:baseline} shows the MSE scores of the 2D network, with and without the STN module, for all the 4 speakers and the 4 extra sessions from speaker 048. In this configuration, we expected no significant benefit from the STN, and this is exactly what we see. The scores we obtained are quite similar accross all speakers and the extra sessions of speaker 048. One notable difference is that the results are consistently worse for the additional sessions of speaker 048 than for her main session. The explanation is that these extra sessions were much shorter than the multi-speaker recordings (3.5 vs. 15 minutes).

In the adaptation experiments we considered the network trained on the data of speaker 048 as our base model. First, we evaluated it on the samples from the other speakers and the additional sessions of the same speaker without any adaptation. These results will be considered as the base results (i.e. $0\%$ relative improvement due to adaptation). Then, we trained this baseline model further on the data from the other speakers and sessions, using three adaptation strategies. First, we allowed the full network to learn on the new data, and the results obtained this way will be considered to have $100\%$ improvement (i.e. the best possible result achievable by adaptation). Next, we froze the weights of the regression network, and allowed only the STN module to adapt. In the final configuration, besides the STN we also allowed the re-adjustment of the weights in the output layer. This layer simply calculates a weighted linear combination of the hidden representations, so allowing it to adapt might significantly decrease the regression error, while it contains only about $1\%$ of all the weights in the network. In all the adaptation experiments we used an initial learning rate that is an order of magnitude smaller than that of the baseline, but otherwise there were no other changes in the training process.

\begin{table*}[t]
    \begin{minipage}{0.3\linewidth}

      \begin{minipage}{0.8\linewidth}
        \vspace{2mm}
      
            \begin{figure}[H]
              \centering
              \includegraphics[width=0.9\linewidth]{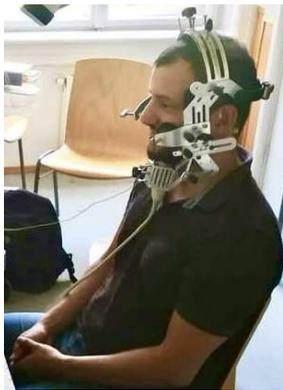}
              \caption{A person has been recorded his speech and tongue movements by the Ultrasound Tongue Imaging device.}
              \vspace{-2mm}
              \label{fig:udi}
            \end{figure}
        \end{minipage}
  \end{minipage}
  \begin{minipage}{0.7\linewidth}
    \caption{MSE rates of the 2D-CNN model of spk048 for the other 3 speakers (top), and for the 4 extra sessions of the same speaker (down), using various adaptation strategies.}
    \centering
    \renewcommand{\arraystretch}{1.1}
    \begin{tabular}{|c||c|c|c|c||c|}
      \hline
        & \multicolumn{5}{c|}{\bf Adaptation method}\\
        \hline
        Speaker~ & None & STN & STN + out & Full & Mean $\theta$\\
        \hline
        \hline
        ~spk049~ & 1.049 & 0.588 \textbf{\scriptsize{(-71\%)}} & 0.517 \textbf{\scriptsize{(-82\%)}} & 0.400 & 0.887 \textbf{\scriptsize{(-25\%)}}\\
        \hline
        ~spk102~ & 1.401 & 0.609 \textbf{\scriptsize{(-78\%)}} & 0.449 \textbf{\scriptsize{(-94\%)}} & 0.389 & 1.015 \textbf{\scriptsize{(-38\%)}}\\
        \hline
        ~spk103~ & 1.322 & 0.552 \textbf{\scriptsize{(-79\%)}} & 0.469 \textbf{\scriptsize{(-88\%)}} & 0.350 &0.909 \textbf{\scriptsize{(-42\%)}}\\
        \hline
        \hline
        ~Average diff.~ & \textbf{-0\%} & \textbf{-76\%} & \textbf{-88\%} & \textbf{-100\%} &\textbf{-35\%}\\
        \hline
        \end{tabular}
        \vspace{1mm}
        
        \centering
        \renewcommand{\arraystretch}{1.1} 
        \begin{tabular}{|c||c|c|c|c||c|}
      \hline
        & \multicolumn{5}{c|}{\bf Adaptation method}\\
        \hline
        Session~ & None & STN & STN + out & Full & Mean $\theta$\\
        \hline
        \hline
        ~ses-2~& 1.131 & 0.646 \textbf{\scriptsize{(-77\%)}} & 0.547 \textbf{\scriptsize{(-93\%)}} & 0.503 & 0.913 \textbf{\scriptsize{(-34\%)}}\\
        \hline
        ~ses-3~& 0.998 & 0.619 \textbf{\scriptsize{(-69\%)}} & 0.485 \textbf{\scriptsize{(-94\%)}} & 0.451 & 0.934 \textbf{\scriptsize{(-11\%)}}\\
        \hline
        ~ses-4~& 1.054 & 0.641 \textbf{\scriptsize{(-70\%)}} & 0.522 \textbf{\scriptsize{(-90\%)}} & 0.468 & 0.908 \textbf{\scriptsize{(-24\%)}}\\
        \hline
        ~ses-5~& 1.174 & 0.604 \textbf{\scriptsize{(-85\%)}} & 0.566 \textbf{\scriptsize{(-91\%)}} & 0.506 & 0.955 \textbf{\scriptsize{(-32\%)}}\\
        \hline
        \hline
        ~Average diff.~ & \textbf{-0\%} & \textbf{-75\%} & \textbf{-92\%} & \textbf{-100\%} & \textbf{-26\%}\\
        \hline
    \end{tabular}
    \vspace{-7mm}
    \label{tab:adapt2D}
  \end{minipage}%
  
\end{table*}
Table~\ref{tab:adapt2D} summarizes the results achieved with the different adaptation strategies for the additional speaker (upper panel) and for the additional sessions of the same speaker (lower panel). As the relative improvements are more informative than the actual MSE values, in parentheses we display the relative error reductions (considering the no-adaptation case as 0\% and full adaptation as 100\%), and in the bottom line we summarize the average improvements. The first column (`None', i.e. no adaptation) clearly shows that the results are unacceptable without any adaptation -- the error rates are actually worse than those for a randomly initialization net without any training. Moreover, the scores in the `Full' adaptation column are pretty si\-mi\-lar to those in the baseline table, so pre-training on speaker 048 seems to be neither beneficial nor detrimental (interestingly even for the extra sessions from the same speaker). By allowing only the STN module to adapt (`STN' column) we can eliminate 75-76\% of the performance gap between the non-adapted and fully adapted models. This is pretty good, considering that the regression network itself is not modified at all, we just simply allow the STN to learn a more optimal affine transformation for the images of the given speaker or session. The improvement is even larger if we also allow the linear output layer of the regression network to adjust its weights to the new speaker or session (`STN~+~out'). In this case, the average error reduction is 88\% for cross-speaker and 92\% for cross-session adaptation. The better cross-session score is reasonable, as one would expect that the differences caused by the misalignment of the device might be easier to compensate by an affine transformation than the inherent anatomical differences between speakers.

\begin{table}[t]
\caption{MSE rates for cross-speaker adaptation, using the 3D-CNN network.} \label{tab:adapt3D}
\centering
\renewcommand{\arraystretch}{1.1} 
\begin{tabular}{|c||c|c|c|c|}
\hline
& \multicolumn{4}{c|}{\bf Adaptation method}\\
\hline
Speaker~ & None & STN & STN + out & Full \\
\hline\hline
~spk049~ & 1.105 & 0.553 \textbf{\scriptsize{(-73\%)}} & 0.497 \textbf{\scriptsize{(-80\%)}} & 0.348\\
\hline
~spk102~ & 1.451 & 0.502 \textbf{\scriptsize{(-84\%)}} & 0.416 \textbf{\scriptsize{(-91\%)}} & 0.315 \\
\hline
~spk103~ & 1.541 & 0.501 \textbf{\scriptsize{(-83\%)}} & 0.418 \textbf{\scriptsize{(-90\%)}} & 0.294 \\
\hline
\hline
~Avg.diff.~ & \textbf{-0\%} & \textbf{-80\%} & \textbf{-87\%} & \textbf{-100\%}\\
\hline
\end{tabular}
\vspace{-5mm}
\end{table}

 As in the baseline evaluation (Table~\ref{tab:baseline}) the STN had no effect on the results, we speculated that maybe it is unnecessary to learn a separate transformation for {\em each image} -- perhaps learning one global $\theta$ for {\em each speaker or session} would be enough. We examined the within-speaker variance of the 6 components of theta for the baseline models, and we indeed found that it was very low, on the order of 0.01, while the between-speaker variance of the $\theta$ vectors was 3-5 times larger. This fact seemed to underpin our conjecture, so we preformed the following simple experiment. Instead of using a unique theta for each image, we simply replaced the STN of the adapted models with the mean $\theta$ vector over the samples of the given speaker/session. The results are shown in the rightmost columns of Table~\ref{tab:adapt2D} ('mean $\theta$'), but the findings are negative: while there is a 25-35\% reduction in the error rates, it is very far from the best achievable. It seems that even though the variance of $\theta$ is small, the minor nuances in the learned transformations per image play an important role in the regression accuracy.
 
 Finally, we extended our experiments to 3D input blocks consisting of 25 subsequent images, using a 3D-CNN for the regression task and a 2D-CNN for the STN module (cf. Section~\ref{Conf3D}). 
 In this case, the baseline model yielded an MSE of 0.275 and 0.278 with and without the STN on the data set of speaker 048. As speaker and session adaptation gave very similar results previously, we repeated only the cross-speaker adaptation experiments with this network, and the results are shown in Table~\ref{tab:adapt3D}. While all the error rates are typically lower than they were for the 2D-CNN, the relative improvements with respect to the various adaptation strategies are very similar. 
\section{Conclusions}

Current tongue ultrasound-based SSI systems are sensitive to changing speakers, or even to a slight displacement of the recording device. In this study we examined whether an STN module is able to counterbalance these factors by applying an affine transform on the input image. When applying a simple 2D-CNN for spectral estimation, we found that allowing only the adaptation of the STN module can reduce the error rate by about 75\%, while allowing also the linear output layer to adapt can compensate for 88-92\% of the error. We also extended the experiments to 3D input blocks, and we observed similar tendencies, although the improvement was somewhat smaller. Considering, however, that the STN module and the output layer altogether contain only 10\% of the weights of the full network, and that the improvements were consistent for cross-session and cross-speaker set-ups, the proposed method might allow for a quicker adaptation of the UTI-processing workflow. 
In the future we plan to run experiments with a 3D localization network and with smaller amounts of adaptation material.

\section{Acknowledgements}

The authors would like to express their gratitude to the following institutions and funding sources for their valuable support in conducting this research. This work was supported by the European Union project RRF-2.3.1-21-2022-00004 within the framework of the Artificial Intelligence National Laboratory and project TKP2021-NVA-09, implemented with the support provided by the Ministry of Innovation and Technology of Hungary from the National Research, Development, and Innovation Fund, financed under the TKP2021-NVA funding scheme and FK 142163 grant. T.G.Cs.~was supported by the Bolyai János Research Fellowship of the Hungarian Academy of Sciences and by the ÚNKP-22-5-BME-316 grant. The RTX A5000 GPU card used was donated by the NVIDIA Corporation.




\bibliographystyle{IEEEtran}

\bibliography{Interspeech2023.bib}

\end{document}